\pdfoutput=1
\documentclass[twocolumn,a4paper,conference]{IEEEtran}
\usepackage[T1]{fontenc}
\usepackage[utf8]{inputenc}

\usepackage{ifthen}

\usepackage{csquotes}
\usepackage{amsmath}
\usepackage{amssymb}
\DeclareMathOperator{\E}{E}

\usepackage{mathtools}

\DeclarePairedDelimiter{\abs}{\lvert}{\rvert}
\DeclarePairedDelimiter{\norm}{\lVert}{\rVert}
\DeclarePairedDelimiter{\parens}{\lparen}{\rparen}
\DeclarePairedDelimiter{\bracks}{\lbrack}{\rbrack}
\DeclareMathOperator{\Real}{Re}

\usepackage{array}

\usepackage{url}
\usepackage[expansion]{microtype}

\input{glyphtounicode.tex}
\input{glyphtounicode-cmr.tex}
\usepackage{hyperref}
\hypersetup{pdfauthor={Laas, Tobias; Nossek, Josef A.; Xu, Wen},pdftitle={Limits of Transmit and Receive Array Gain in Massive MIMO}}
\usepackage{doi}

\usepackage{bm}
\newcommand{\matt}[1]{\bm{#1}}
\newcommand{\vect}[1]{\bm{#1}}

\usepackage{siunitx}

\usepackage{etoolbox}
\newbool{finalsubmission}
\setbool{finalsubmission}{true}
\newbool{pdffigures}
\setbool{pdffigures}{false}
\ifbool{finalsubmission}{\setbool{pdffigures}{true}}{}
\ifbool{CLASSOPTIONdraftcls}{\setbool{pdffigures}{true}}{}

\usepackage{tikz}
\ifthenelse{\boolean{finalsubmission}}{}{
\usepackage[europeanresistors,straightvoltages]{circuitikz}
\usetikzlibrary{scopes}
\ctikzset{bipoles/thickness=1}

\sisetup{math-ohm  =\Omega,output-complex-root=\ensuremath{j}}

\newlength{\dotsclength}
\setlength{\dotsclength}{\widthof{\footnotesize$\dotsc$}}

\usetikzlibrary{decorations.markings,chains}
\tikzset{threedots/.style={decorate,decoration={markings,mark=between			positions 0.25 and 0.75 step 0.25 with {\node[circle,draw=black,fill=black,inner sep=0pt,minimum	size=1pt,thin]{};}}}}

\usepackage{pgfplots}
\pgfplotsset{compat=1.14}
\pgfplotsset{every axis/.append style={line width=0.75pt,tick style={line width=0.25pt},grid style={line width=0.5pt}}}

\usetikzlibrary{arrows.meta,calc,positioning}

\usepackage[outline]{contour}
\contourlength{0.8pt}

\tikzset{double_arrow/.style={-{Triangle[length=3.9pt,white,width=5pt]},double distance=5pt,thick,postaction={decorate,decoration={markings,mark=at position 1 with {\arrow[scale=2.5,thin]{Straight Barb}}}}
}}

\tikzset{vh path/.style={to path={|- (\tikztotarget)}},hv path/.style={to path={-| (\tikztotarget)}}}

\tikzset{adder/.style={circle,minimum size=.25cm,inner sep=0pt,draw=black,very thick,execute at end node={$\textbf{+}$}}}

\tikzset{filter/.style={rectangle,inner sep=2pt,minimum height=0.6cm,draw=black,very thick}}

\usetikzlibrary{backgrounds}
\tikzstyle{thicker}=[line width=1pt]

\usetikzlibrary{external}
\tikzexternalize[prefix=external/]
}

\DeclareSIUnit\bpcu{bpcu}
\DeclareSIUnit\decibelwatt{dBW}

\usepackage{textcomp}

\makeatletter
\let\MYcaption\@makecaption
\makeatother

\usepackage[font=footnotesize]{subcaption}

\makeatletter
\let\@makecaption\MYcaption
\makeatother

\begin{document}
\title{Limits of Transmit and Receive Array Gain in Massive MIMO}

\author{\IEEEauthorblockN{Tobias Laas\IEEEauthorrefmark{1}\IEEEauthorrefmark{2}, 
		Josef A. Nossek\IEEEauthorrefmark{2}\IEEEauthorrefmark{3}, Wen Xu\IEEEauthorrefmark{1}}
	\IEEEauthorblockA{\IEEEauthorrefmark{1}German Research Center, Huawei Technologies Duesseldorf GmbH, Munich, Germany}
	\IEEEauthorblockA{\IEEEauthorrefmark{2}Department of Electrical and Computer Engineering, Technical University of Munich, Munich, Germany}
	\IEEEauthorblockA{\IEEEauthorrefmark{3}Department of Teleinformatics Engineering, Federal University of Ceará, Fortaleza, Brazil}
	\IEEEauthorblockA{Emails: \{tobias.laas \& josef.a.nossek\}@tum.de, wen.xu@ieee.org}}

\maketitle
\begin{tikzpicture}[remember picture,overlay]
\node[yshift=2cm] at (current page.south){\parbox{\textwidth}{\footnotesize \textcopyright\ 2020 IEEE. Personal use of this material is permitted. Permission from IEEE must be obtained for all other uses, in any current or future media, including reprinting/republishing this material for advertising or promotional purposes, creating new collective works, for resale or redistribution to servers or lists, or reuse of any copyrighted component of this work in other works.}};
\node[yshift=-1cm] at (current page.north){\parbox{\textwidth}{\footnotesize This is the accepted version of the following article: 
T.~Laas, J.~A. Nossek, and W.~Xu, ``Limits of transmit and receive array gain in massive {MIMO},'' in \emph{Proc. IEEE Wireless Commun. Netw. Conf. (WCNC)}, May 2020, \doi{10.1109/WCNC45663.2020.9120590}.}};
\end{tikzpicture}%
\begin{abstract}
In this paper, we consider the transmit and receive antenna array gain of massive MIMO systems. In particular, we look at their dependence on the number of antennas in the array, and the antenna spacing for uniform linear and uniform circular arrays. It is known that the transmit array gain saturates at a certain antenna spacing, but the receive array gain had not been considered. With our physically consistent analysis based on the Multiport Communication Theory, we show that the receive array gain does not saturate, but that there is a peak at a certain antenna spacing when there is no decoupling network at the receiver. As implementing a decoupling network for massive MIMO would be almost impossible, this is a reasonable assumption. Furthermore, we analyze how the array gain changes depending on the antenna spacing and the size of the antenna array and derive design recommendations.
\end{abstract}%
\section{Introduction}
Massive MIMO is an important building block of future wireless systems, as, depending on the scenario, a larger number of base station antennas is believed to increase the achievable transmit and receive array gain, i.e., it allows for a larger SNR at the same transmit power, a lower bit error ratio by exploiting diversity, or to serve more mobiles at the same time. Indeed, the seminal paper~\cite{Marzetta2010} that introduced massive MIMO is based on the assumption that there is an unlimited number of base station antennas. However for realistic systems, does increasing the number of base station antennas always improve performance?

In~\cite{IvrlacWSA2016}, it has been shown for a uniform circular array (UCA) at the base station transmitting to a mobile over a line-of-sight (LOS) channel (without reflections) that in general the minimum energy per bit $E_\mathrm{b,min}$, which is inversely proportional to the transmit array gain, decreases as the number of antennas at the base station increases, but at a certain number of base station antennas, $E_\mathrm{b,min}$ saturates. The analysis is based on the Multiport Communication Theory~\cite{IvrlacNossekTowardaTheory,IvrlacNossekMultiportCommTheory}, which is in turn based on circuit theory and ensures that the analysis is physically consistent.

One contribution of this paper is to extend the analysis to the receive array gain. We also want to extend the analysis to antenna arrays, where the antenna separation is fixed rather than the array size. Note that transmit and receive array gain are different, unless the noise at the receiver fulfills certain properties~\cite{IvrlacNossekTowardaTheory}, as we define array gain as the ratio of SNRs instead of powers. Another contribution of this paper is to look at how the array gain changes if both antenna spacing and array size vary and derive design recommendations. The influence of mutual coupling on transmit array gain has already been investigated early~\cite{Uzkov}. Experimental results were provided in~\cite{Yaghjian}, but only for small arrays and without investigating the difference between transmit and receive array gain.

\textit{Notation}: lowercase bold letters denote vectors, uppercase bold letters matrices. $a_m$ denotes the $m$th element of $\vect a$. $\matt A^T, \matt A^\ast$ and $\matt A^H$, correspond to the transpose, the complex conjugate and the Hermitian. $\vect 0$ and $\matt I$ denote zero vector and identity matrix. $\mathcal{N}_\mathbb{C}(\vect \mu, \matt R)$ denotes a circularly-symmetric complex Gaussian distribution with mean $\vect \mu$ and covariance $\matt R$.  $\E[ X ]$ denotes the expectation of the random variable $X$.

\section{Theory}
\label{sec:theory}
Similarly to~\cite{IvrlacWSA2016}, we consider a multi-antenna transmitter and a single antenna receiver, and do not use the unilateral approximation because the currents in the receive antennas do influence the transmit antennas, so that the near field is important to the analysis. See~\cite{IvrlacNossekTowardaTheory,IvrlacNossekMultiportCommTheory}, for more details on the unilateral approximation. In addition, we also consider the reverse link with a single antenna transmitter and a multi-antenna receiver.

\begin{figure*}[t]
	\centering
	\ifbool{pdffigures}{%
		\includegraphics{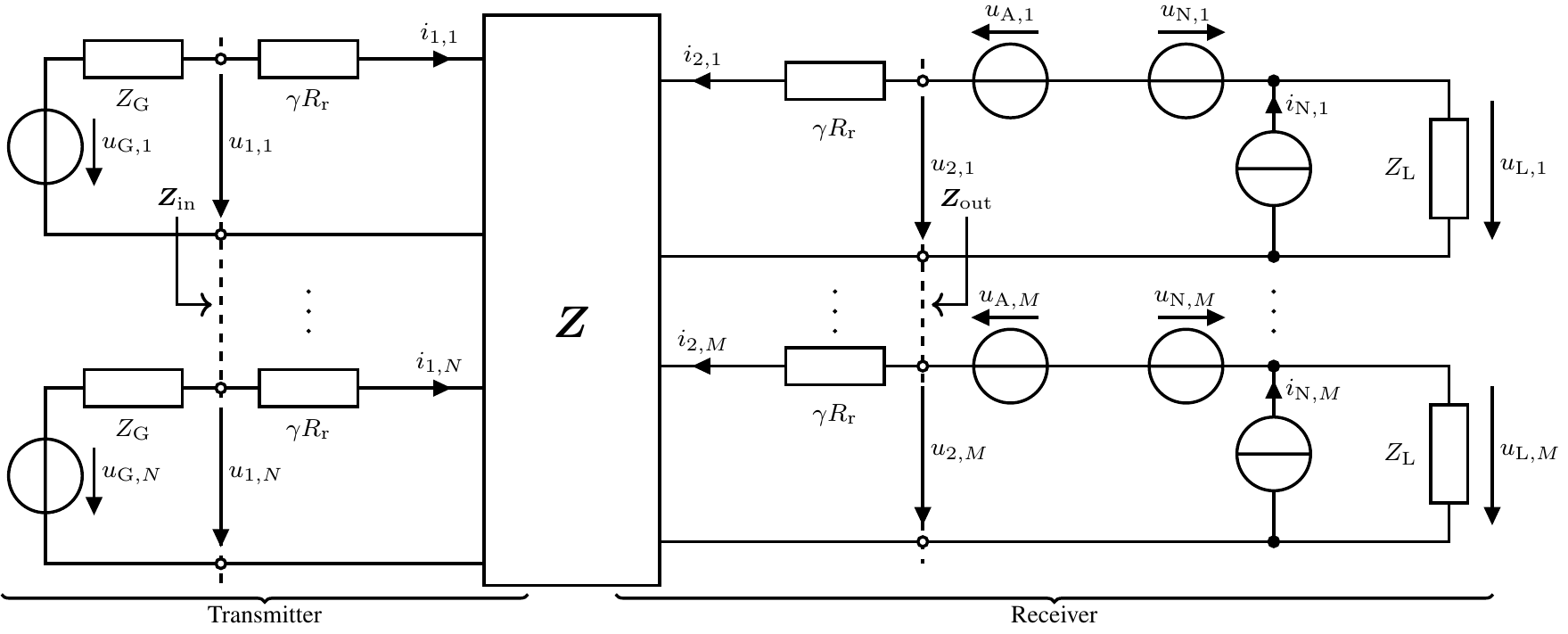}%
	}{%
		\tikzsetnextfilename{impedancenetworkDL}%
		\input{impedancenetworkDL}%
	}%
	\caption{Circuit model (modified from~\cite{WSAjournal}).}%
	\label{fig:circuitmodelDL}%
\end{figure*}%

Consider the circuit model for a setup with $N$ transmit and $M$ receive antennas, see Fig.~\ref{fig:circuitmodelDL}. The lossless decoupling and (impedance) matching networks (DMNs) are omitted because in massive MIMO systems, they would be almost impossible to implement.

The transmit amplifiers are modeled as linear amplifiers. Let $Z_\mathrm{G}$ be their internal resistance and $\vect u_\mathrm{G}$ their open load voltage. Power matching is employed at the transmitter, i.e.,
\begin{equation}
Z_\mathrm{G} = Z_\mathrm{A}^\ast,\quad R_\mathrm{r} \coloneqq \Real(Z_\mathrm{A}),\quad R_\mathrm{G} = \Real\parens{Z_\mathrm{G}},
\end{equation}
where $Z_\mathrm{A}$ is the self-impedance of the antennas and $R_\mathrm{r}$ their radiation resistance. Let $Z_\mathrm{L}$ be the input impedance of the low noise amplifier (LNA) in each RF chain, $\vect u_\mathrm{L}$ the load voltage and
\begin{equation}
R_\mathrm{L} = \Real\parens{Z_\mathrm{L}}.
\end{equation}

The impedance matrix $\matt Z$ can be partitioned into four blocks~\cite{IvrlacNossekTowardaTheory}
\begin{equation}
\matt Z = \begin{bmatrix} \matt Z_{11} & \matt Z_{12}\\
\matt Z_{21} & \matt Z_{22} \end{bmatrix} \in \mathbb{C}^{(N+M)\times(N+M)},
\end{equation}
the transmit and receive impedance matrices $\matt Z_{11} \in \mathbb{C}^{N\times N}$, and $\matt Z_{22} \in \mathbb{C}^{M \times M}$ and the mutual impedance matrices $\matt Z_{21} \in \mathbb{C}^{M \times N}$ and $\matt Z_{12} \in \mathbb{C}^{N \times M}$, where $\matt Z=\matt Z^T$ due to reciprocity. Let
\begin{equation}
\matt Z_{11,\mathrm{r}}=\matt Z_{11} + \gamma R_\mathrm{r} \matt I, \quad \matt Z_{22,\mathrm{r}}=\matt Z_{22} +\gamma R_\mathrm{r} \matt I,
\end{equation}
where the dissipation resistance $\gamma R_\mathrm{r}$, which is connected in series, is used to model the losses in the antennas, see Fig.~\ref{fig:circuitmodelDL}. The impedance matrices seen at the input and the output are
\begin{align}
\matt Z_\mathrm{in} = &\matt Z_{11,\mathrm{r}}
- \matt Z_{12} \parens{Z_\mathrm{L}\matt I + \matt Z_{{22},\mathrm{r}}}^{-1} \matt Z_{21},
\\
\matt Z_\mathrm{out} = &\matt Z_{22,\mathrm{r}} - \matt Z_{21} \parens{ Z_\mathrm{G} \matt I + \matt Z_{11,\mathrm{r}}}^{-1} \matt Z_{12}.
\end{align}
The physical model (\cite{IvrlacWSA2016,WSAjournal} combined) is
\begin{gather}
\label{equ:physicalmodel}
\begin{split}
\vect u_\mathrm{L} = \matt D \vect u_\mathrm{G} + \sqrt{R_\mathrm{L}} \vect \eta, \quad \vect \eta \sim \mathcal{N}_\mathbb{C}( \vect 0\,\si{\sqrt{\watt}}, \matt R_\eta),\\ \quad P_\mathrm{T} = \frac{\E[\vect u_\mathrm{G}^H \matt B \vect u_\mathrm{G}]}{R_\mathrm{G}},
\end{split}\\
\matt R_\eta = \frac{\abs{Z_\mathrm{L}}^2}{R_\mathrm{L}} ( \matt Z_\mathrm{out} + Z_\mathrm{L} \matt I )^{-1} \matt Q ( \matt Z_\mathrm{out} + Z_\mathrm{L} \matt I )^{-H}\\
\matt B = R_\mathrm{G} \parens*{\matt Z_\mathrm{in} + Z_\mathrm{G} \matt I}^{-H} \Real(\matt Z_\mathrm{in}) \parens*{\matt Z_\mathrm{in} + Z_\mathrm{G} \matt I }^{-1},\\
\matt D = Z_\mathrm{L} (\matt Z_{22} + Z_\mathrm{L} \matt I )^{-1} \matt Z_{21} ( \matt Z_\mathrm{in}+ Z_\mathrm{G} \matt I )^{-1},
\end{gather}
where $\matt D$ describes the noiseless relation between $\vect u_\mathrm{G}$ and $\vect u_\mathrm{L}$, $\vect \eta$ describes the noise, $\matt B$ is the power-coupling matrix, $P_\mathrm{T}$ is the transmit power and $\matt Q$ is a noise covariance matrix. $\matt Q$ comes from intrinsic noise sources $\vect u_\mathrm{N}$ and $\vect i_\mathrm{N}$ and the antenna noise $\vect u_\mathrm{A}$, and is defined as~\cite{WSAjournal}
\begin{gather}
\matt Q = \sigma_u^2 \matt I + \sigma_i^2 \matt Z_\mathrm{out} \matt Z_\mathrm{out}^H - 2 \sigma_u \sigma_i \Real\parens{\rho^\ast \matt Z_\mathrm{out}}+\matt R_\mathrm{A},\\
\matt R_\mathrm{A}= 4 k_\mathrm{B} T_\mathrm{A} \Delta f \Real\parens{\matt Z_\mathrm{out}},
\end{gather}
where $k_\mathrm{B}$ is the Boltzmann constant, $\Delta f$ is the noise bandwidth, $T_\mathrm{A}$ is the noise temperature of the antennas and $\sigma_u, \sigma_i, \rho$ describe the zero-mean circularly-symmetric complex Gaussian distribution of $\vect u_\mathrm{N}$ and $\vect i_\mathrm{N}$ in the equivalent two-port model at the receiver, similarly to~\cite{WSAjournal}.

The corresponding information-theoretic model (\cite{IvrlacWSA2016,WSAjournal} combined) is
\begin{gather}
\label{equ:infomodel}
\begin{split}
\vect y &= \matt H \vect x + \vect \vartheta, \quad \vect \vartheta \sim \mathcal{N}_\mathbb{C}(\vect 0\,\sqrt{\mathrm{W}}, \sigma_\vartheta^2 \matt I), \quad P_\mathrm{T} = \E\bracks{\norm{\vect x}_2^2},\\
\matt H 
&= \sigma_\vartheta \frac{\sqrt{R_\mathrm{G}}}{\sqrt{R_\mathrm{L}}} \matt R_\eta^{-1/2} \matt D \matt B^{-H/2},
\end{split}\raisetag{20pt}
\end{gather}
where $\vect y$ is the received signal, $\vect x$ is the transmitted signal, $\matt H$ is the information-theoretic channel and $\vect \vartheta$ is additive white Gaussian noise.

By choosing specific $\matt B^{1/2}$ and $\matt R_\eta^{1/2}$ to transform between the physical model \eqref{equ:physicalmodel} and the model \eqref{equ:infomodel} analogously to~\cite{WSAjournal},
\begin{align}
\label{equ:defbsqrt}
\begin{split}
\matt B^{1/2} &= \sqrt{R_\mathrm{G}}\parens*{\matt Z_\mathrm{in} + Z_\mathrm{G} \matt I}^{-H} \Real(\matt Z_\mathrm{in})^{1/2}\\&\mathrm{s.t}\quad \Real(\matt Z_\mathrm{in})=\Real(\matt Z_\mathrm{in})^{1/2}\Real(\matt Z_\mathrm{in})^{1/2},
\end{split}\\
\matt R_\eta^{1/2} &= \frac{Z_\mathrm{L}}{\sqrt{R_\mathrm{L}}} \parens*{\matt Z_\mathrm{out} + Z_\mathrm{L} \matt I}^{-1} \matt{Q}^{1/2},
\end{align}
we can write 
\begin{gather}
\matt H = \sigma_\vartheta \matt Q^{-1/2} \matt Z_{21,\mathrm{eff}} \Real(\matt Z_\mathrm{in})^{-1/2},\\
\label{equ:defz21eff}
\begin{split}
\matt Z_{21,\mathrm{eff}} = &\matt Z_{21}\\& - \matt Z_{21} \parens{Z_\mathrm{G}\matt I + \matt Z_{11,\mathrm{r}}}^{-1} \matt Z_{12} \parens{Z_\mathrm{L} \matt I + \matt Z_{22,\mathrm{r}}}^{-1} \matt Z_{21},
\end{split}\raisetag{28pt}
\end{gather}
where $\matt Z_{21,\mathrm{eff}}$ is the effective mutual impedance matrix between transmitter and receiver.

\subsection{Receive Array Gain}
For the receive array gain, we consider an uplink scenario with a mobile with one antenna transmitting to a base station with $N_\mathrm{BS}$ antennas, i.e., $N=1, M=N_\mathrm{BS}$. This implies that $\matt H$ becomes a vector $\vect h \coloneqq \matt H$ and similarly $\vect z_{21,\mathrm{eff}} \coloneqq \matt Z_{21,\mathrm{eff}}$ and $Z_\mathrm{in} \coloneqq \matt Z_\mathrm{in}$.
The receive array gain is defined as~\cite{IvrlacNossekTowardaTheory}
\begin{equation}
\label{equ:defarxgain}
A_\mathrm{Rx}\coloneqq\left.\frac{\max \mathrm{SNR}}{\mathrm{SNR}\rvert_{M=1,\gamma=0}}\right\rvert_{P_\mathrm{T}=\mathrm{const.}},
\end{equation}
where
\begin{equation}
\max \mathrm{SNR} = \frac{\norm{\vect h}_2^2}{\sigma_\vartheta^2} P_\mathrm{T} = \frac{\vect z_{21,\mathrm{eff}}^H \matt Q^{-1} \vect z_{21,\mathrm{eff}}}{\Real\parens{Z_\mathrm{in}}} P_\mathrm{T}
\end{equation}
is obtained by use of a matched filter at the receiver and the $\mathrm{SNR}$ for $M=1$ lossless receive antennas is obtained in a similar way. Then
\begin{equation}
\label{equ:defarx}
A_\mathrm{Rx}=\frac{\Real\parens{Z_{\mathrm{in},0}}}{\Real\parens{Z_{\mathrm{in}}}}\frac{\vect z_{21,\mathrm{eff}}^H \matt Q^{-1} \vect z_{21,\mathrm{eff}} \sigma_{q,0}^2}{\abs{z_{21,\mathrm{eff},0}}^2},
\end{equation}
where
\begin{equation}
\label{equ:defaqzero}
\begin{split}
Z_{\mathrm{in},0} \coloneqq \matt Z_{\mathrm{in}}\rvert_{M=1,\gamma=0},\qquad \sigma_{q,0}^2 \coloneqq \matt Q \rvert_{M=1,\gamma=0},\\ z_{21,\mathrm{eff},0} \coloneqq \vect z_{21,\mathrm{eff}} \rvert_{M=1,\gamma=0}.
\end{split}
\end{equation}

\subsection{Transmit Array Gain}
\begin{figure}[t]
	\begin{minipage}[b]{\linewidth}
		\centering
		\ifbool{pdffigures}{%
			\includegraphics{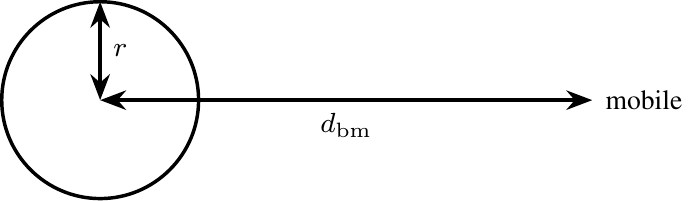}%
		}{%
			\tikzsetnextfilename{scenarioa}%
			\input{scenarioa}%
		}%
		\subcaption{UCA with fixed radius.}%
		\label{fig:secnarioUCA}
	\end{minipage}\\
	\begin{minipage}[b]{\linewidth}
		\ifbool{pdffigures}{%
			\includegraphics{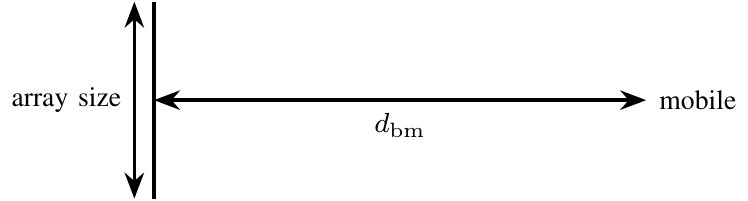}%
		}{%
			\tikzsetnextfilename{scenariob}%
			\input{scenariob}%
		}%
		\subcaption{ULA transmitting into frontfire direction.}%
		\label{fig:secnarioULA}
	\end{minipage}%
	\caption{Scenarios.}%
	\label{fig:scenarios}%
\end{figure}%
For the transmit array gain, we consider a downlink scenario with a base station with $N_\mathrm{BS}$ antennas transmitting to a mobile with one antenna, i.e., $N=N_\mathrm{BS},M=1$. Here, $\vect h \coloneqq \matt H^H$ and $\vect z_{21,\mathrm{eff}} \coloneqq \matt Z_{21,\mathrm{eff}}^H$ to make them column vectors and $\matt Q$ becomes a scalar $\sigma_q^2$.
The transmit array gain is defined as~\cite{IvrlacNossekTowardaTheory}
\begin{equation}
\label{equ:defatxgain}
A_\mathrm{Tx}\coloneqq\left.\frac{\max \mathrm{SNR}}{\mathrm{SNR}\rvert_{N=1,\gamma=0}}\right\rvert_{P_\mathrm{T}=\mathrm{const.}},
\end{equation}
where
\begin{equation}
\max \mathrm{SNR} = \frac{\norm{\vect h}_2^2}{\sigma_\vartheta^2} P_\mathrm{T} = \frac{\vect z_{21,\mathrm{eff}}^H \Real\parens{\matt Z_\mathrm{in}}^{-1} \vect z_{21,\mathrm{eff}}}{\sigma_q^2} P_\mathrm{T}
\end{equation}
is obtained by using a matched filter at the transmitter and the $\mathrm{SNR}$ for $N=1$ lossless transmit antennas is obtained in a similar way. Then,
\begin{equation}
\label{equ:defatx}
A_\mathrm{Tx}=\frac{\Real\parens{Z_{\mathrm{in},0}}}{\sigma_q^2}\frac{\vect z_{21,\mathrm{eff}}^H \Real\parens{\matt Z_\mathrm{in}}^{-1} \vect z_{21,\mathrm{eff}} \sigma_{q,0}^2}{\abs{z_{21,\mathrm{eff},0}}^2},
\end{equation}
where $Z_\mathrm{in,0}, \sigma_{q,0}^2$ and $z_{21,\mathrm{eff},0}$ are defined as in \eqref{equ:defaqzero}, but for $N=1,\gamma=0$.

Considering the transmit array gain is equivalent to considering the minimum transmitted energy per bit $E_\mathrm{b,min}$ as in~\cite{IvrlacWSA2016}. This can be shown as follows: by adding the losses in the antennas to the model in \cite{IvrlacWSA2016} and for the more general noise distribution assumed in this paper, 
\begin{equation}
E_\mathrm{b,min} = \frac{\sigma_q^2 \ln 2}{\Delta f\vect z_{21,\mathrm{eff}}^H \Real\parens{\matt Z_\mathrm{in}}^{-1} \vect z_{21,\mathrm{eff}}}.
\end{equation}
This means
\begin{equation}
A_\mathrm{Tx} \propto E_\mathrm{b,min}^{-1}.
\end{equation}
Note that transmitting with $E_\mathrm{b,min}$ leads to the well-known minimum received energy per bit $\sigma_\vartheta^2 \ln\parens{2}/(\Delta f)$, see~\cite{VerduWidebandRegime}.

\subsection{Channel}
In the following we assume that all antennas at the base station and the mobile are parallel infinitely thin but lossless $\lambda/2$-dipoles in series with the dissipation resistance $\gamma R_\mathrm{r}$. Then for a line of sight channel, the entries of $\matt Z$ can be computed according to the analytical formulas using the sinusoidal current approximation~\cite{IvrlacWSA2016,SchelkunoffFriis} because they are canonical minimum scattering antennas~\cite{MSant,CMSAnt}.

For a receiver located in the far field in direction $(\theta,\varphi)$, where $\theta$ is the zenith angle and $\varphi$ the azimuth angle, 
\begin{equation}
\vect z_{21}  = R_\mathrm{r} \vect a(\theta,\varphi),
\end{equation}
where $\vect a$ is the steering vector, i.e.,

\begin{equation}
a_i(\theta,\varphi) = e^{j \frac{2\pi}{\lambda} \vect r_i^T \vect r}, \quad \vect r = \begin{bmatrix}
\cos(\varphi) \sin(\theta)\\
\sin(\varphi) \sin(\theta)\\
\cos(\theta)
\end{bmatrix}
\end{equation}
and $\vect r_i$ is the position vector of the $i$-th antenna. We choose the coordinate system such that the origin coincides with the center of the array. The $\lambda/2$-dipoles are oriented parallel to $\theta = 0$, the ULAs are oriented such that they lie in $\varphi=\pi/2$ and the UCAs are oriented such that one antenna lies in $\varphi=0$.

Extending the consideration to the far field,
\begin{equation}
\begin{split}
Z_{\mathrm{in},0} \to Z_\mathrm{A}, \quad \vect z_{21,\mathrm{eff}} \to \vect z_{21}, \quad \matt Z_\mathrm{in} \to \matt Z_{11,\mathrm{r}},\\
\sigma_{q}^2 \to \sigma_{q,0}^2, \quad z_{21,\mathrm{eff},0} \to z_{21,0}, \quad \matt Z_\mathrm{out} \to \matt Z_{22,\mathrm{r}},
\end{split}
\end{equation}
where $\vect z_{21}$ and $z_{21,0}$ are defined analogously to $\vect z_{21,\mathrm{eff}}$ and $z_{21,\mathrm{eff},0}$. That means, $A_\mathrm{Tx}$ in the far field is~\cite{IvrlacNossekTowardaTheory}
\begin{equation}
\label{equ:atxff}
A_\mathrm{Tx}=R_\mathrm{r}\frac{\vect z_{21}^H \Real\parens{\matt Z_{11,\mathrm{r}}}^{-1} \vect z_{21}}{\abs{z_{21,0}}^2}.
\end{equation}
Different arrays vary in $\Real\parens{\matt Z_{11,\mathrm{r}}}$ and in $\vect a(\theta,\varphi)$. The former is Toeplitz for ULAs and circulant for UCAs.

\section{Numerical Results}
\begin{table*}[!t]
	\renewcommand{\arraystretch}{1.3}
	\caption{Overview of the distances between the base station and the mobile at the different frequencies.}%
	\label{tab:frequencies}%
	\centering%
	\begin{tabular}{|c|c|c|c|}
		\hline
$d_\mathrm{bm}$ & @ $f_c = \SI{680.5}{\mega\hertz}$ (n71 uplink)& @ $f_c = \SI{3.55}{\giga\hertz}$ (n78) & @ $f_c = \SI{27.925}{\giga\hertz}$ (n261)\\
\hline 
$10^2\lambda$ & $\SI{44.1}{\meter}$ & $\SI{8.44}{\meter}$ & $\SI{1.07}{\meter}$\\
\hline
$10^{2.5}\lambda$ & $\SI{139}{\meter}$ & $\SI{26.7}{\meter}$ & $\SI{3.39}{\meter}$\\
\hline
$10^{3}\lambda$ & $\SI{441}{\meter}$ & $\SI{84.4}{\meter}$ & $\SI{10.7}{\meter}$\\
\hline
$10^{3.5}\lambda$ & $\SI{1.39}{\kilo\meter}$ & $\SI{267}{\meter}$ & $\SI{33.9}{\meter}$\\
\hline
$10^{4}\lambda$ & $\SI{4.41}{\kilo\meter}$ & $\SI{844}{\meter}$ & $\SI{107}{\meter}$\\
\hline
	\end{tabular}%
\end{table*}%

\begin{table}[!t]
	\renewcommand{\arraystretch}{1.3}
	\caption{Overview of 3GPP 38.901 channel model parameters.}%
	\label{tab:chmodelparams}%
	\centering%
	\begin{tabular}{|c|c|c|c|}
		\hline
&RMa & UMa & UMi\\
\hline
Base station altitude & $\SI{35}{\meter}$ & $\SI{25}{\meter}$ & $\SI{10}{\meter}$ \\
\hline
Mobile altitude & $\SI{1.5}{\meter}$ & \multicolumn{2}{c|}{\begin{minipage}[t]{0.9in}between $\SI{1.5}{\meter}$\\ and $\SI{22.5}{\meter}$\end{minipage}}\\[9pt]
\hline 
Minimum horizontal distance& \multicolumn{2}{c|}{$\SI{35}{\meter}$} & $\SI{10}{\meter}$\\
\hline
Inter site distance & $\SI{5000}{\meter}$ & $\SI{500}{\meter}$ & $\SI{200}{\meter}$ \\
\hline
$\Rightarrow$ Minimum distance & $\SI{48.4}{\meter}$ & $\SI{35.1}{\meter}$ & $\SI{10.0}{\meter}$\\
\hline
$\Rightarrow$ Maximum distance & $\SI{2.89}{\kilo\meter}$ & $\SI{290}{\meter}$ & $\SI{116}{\meter}$\\
\hline
	\end{tabular}%
\end{table}%

Consider the distance $d_\mathrm{bm}$ between the base station and the mobile, see Fig.~\ref{fig:scenarios}. In the following section,
\begin{equation}
d_\mathrm{bm} \in \{10^{i/2}\lambda \mid  i=4,\dotsc,8\}.
\end{equation}
Table~\ref{tab:frequencies} shows the value of these distances for the following frequency bands:
\begin{itemize}
	\item The uplink in LTE (5G NR) band 71 (n71), which spans $\SI{663}{\mega\hertz}$ to $\SI{698}{\mega\hertz}$~\cite{3GPP36101-f50,3GPP38101-1-f40}, i.e., $f_\mathrm{c} = \SI{680.5}{\mega\hertz}$, which is one of the lowest frequency bands currently supported for mobile broadband.
	\item Band n78, from $\SI{3.3}{\giga\hertz}$ to $\SI{3.8}{\giga\hertz}$~\cite{3GPP38101-1-f40}, i.e., $f_\mathrm{c} = \SI{3.55}{\giga\hertz}$.
	\item 5G NR mmWave band n261, which spans $\SI{27.5}{\giga\hertz}$ to $\SI{28.35}{\giga\hertz}$~\cite{3GPP38101-2-f40}, i.e. $f_\mathrm{c} = \SI{27.925}{\giga\hertz}$.
\end{itemize}
All three bands are among the first for 5G NR deployment. 

Let us compare these distances to the ones in the 3GPP 38.901 channel model~\cite{3GPP38901-1410}. For band n71, consider the rural macro (RMa), for band n78 the urban macro (UMa) and for band n261 the urban micro (UMi) scenario. The base station altitude, mobile height, minimum 2D distance and inter site distance are as shown in Table~\ref{tab:chmodelparams}. According to these parameters, a $d_\mathrm{bm}$ between $100 \lambda$ and $10000 \lambda$ almost entirely covers the various scenarios. In the following, the various frequency bands will be analyzed jointly with all distances normalized to $\lambda$.

\begin{figure}[t]
	\centering
	\ifbool{pdffigures}{%
		\includegraphics{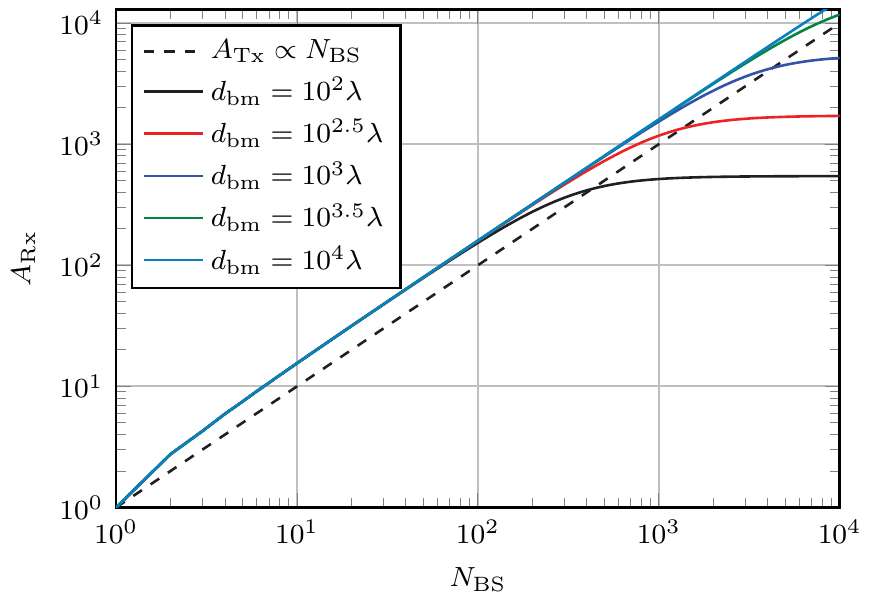}%
	}{%
		\tikzsetnextfilename{rxtxlimULAdstfixRx}%
		\input{rxtxlimULA_dstfix_Rx}%
	}%
	\caption{$A_\mathrm{Rx}$ for a ULA in frontfire with fixed antenna separation $d=\lambda/2$.}%
	\label{fig:rxtxlimULAdstfixRx}
\end{figure}%
\begin{figure}[t]
	\centering
	\ifbool{pdffigures}{%
		\includegraphics{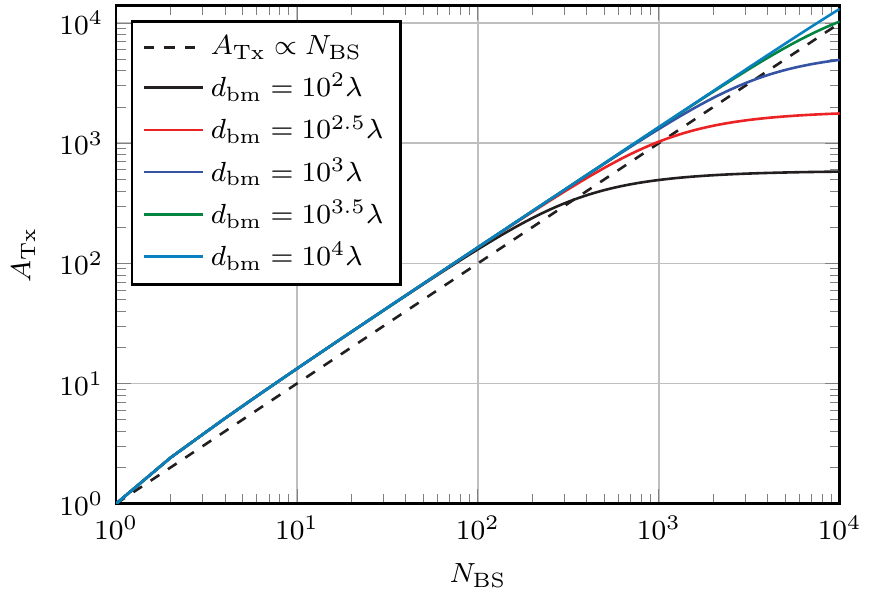}%
	}{%
		\tikzsetnextfilename{rxtxlimULAdstfixTx}%
		\input{rxtxlimULA_dstfix_Tx}%
	}%
	\caption{$A_\mathrm{Tx}$ for a ULA in frontfire with fixed antenna separation $d=\lambda/2$.}%
	\label{fig:rxtxlimULAdstfixTx}
\end{figure}%
In this section, we consider $A_\mathrm{Rx}$ and $A_\mathrm{Tx}$ according to \eqref{equ:defarx} and \eqref{equ:defatx}. We use the measured noise parameters from~\cite[Tables IV \& VI]{LehmeyerJournalPrinted}, i.e., in particular $Z_\mathrm{L} = \SI{186-j31.6}{\ohm}$, and assume that the loss factor $\gamma = 10^{-3}$.

\subsection{Fixed Distance}
Consider a ULA with a fixed antenna separation $d=\lambda/2$ and between $N_\mathrm{BS}=1$ and $N_\mathrm{BS}=10000$ base station antennas transmitting into the frontfire direction, see Fig.~\ref{fig:secnarioULA}. $A_\mathrm{Rx}$ and $A_\mathrm{Tx}$ for this scenario are shown in Figs.~\ref{fig:rxtxlimULAdstfixRx} and \ref{fig:rxtxlimULAdstfixTx}. They are almost identical here. 
According to conventionally modeled systems, which neglect mutual coupling, we expect
\begin{equation}
\label{equ:convgain}
A_\mathrm{Rx} = N_\mathrm{BS}, \qquad A_\mathrm{Tx} = N_\mathrm{BS}.
\end{equation}
However, when $N_\mathrm{BS}$ increases from $1$, the array gain becomes slightly larger than $N_\mathrm{BS}$. Furthermore for a larger $N_\mathrm{BS}$, the array gain starts to saturate, with saturation occurring at a lower $N_\mathrm{BS}$ the smaller the $d_\mathrm{bm}$. For $d_\mathrm{bm}=100\lambda$, the onset of saturation starts at values as low as $N_\mathrm{BS}=100$ to $300$, corresponding to an array size between $49.5\lambda$ and $149.5\lambda$, whereas for $d_\mathrm{bm}=10^{3.5}\lambda$ only the very start of the saturation near $N_\mathrm{BS}=10^4$ can be seen and for $d_\mathrm{bm}=10^4$, the array gain only saturates for an even larger value of $N_\mathrm{BS}$. When we compare the array size at which saturation starts to $d_\mathrm{bm}$, we can see that the array size there is on the same order of magnitude as $d_\mathrm{bm}$. This indicates that saturation occurs as the additional antennas' path-loss increases so they contribute less. For typical cellular systems, whose array size is significantly smaller than the distance between the base station and the mobile, this is not a practical limitation.

\subsection{Fixed Radius}
\label{sec:mmimofixedrad}
\begin{figure}[t]
	\centering
	\ifbool{pdffigures}{%
		\includegraphics{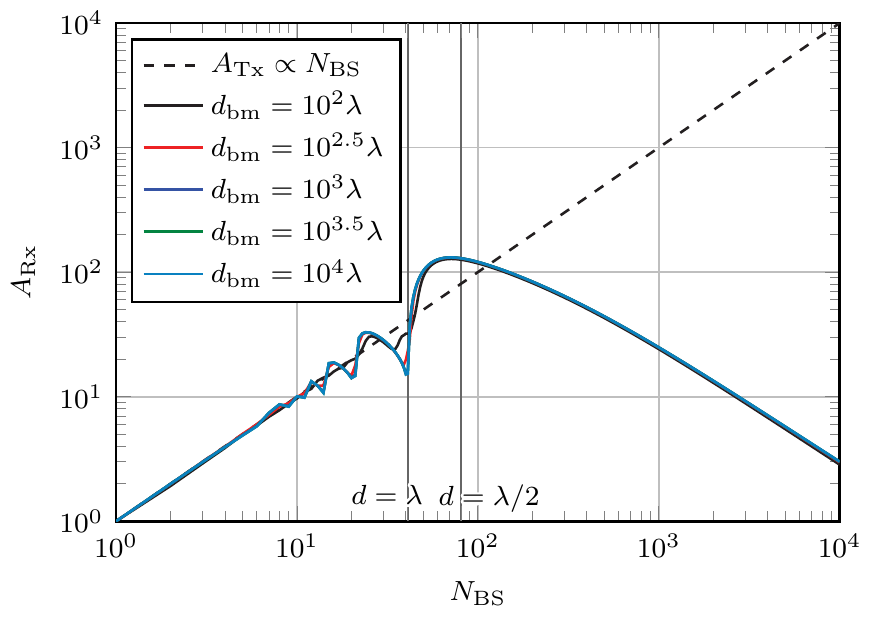}%
	}{%
		\tikzsetnextfilename{rxtxlimULAradfixRx}%
		\input{rxtxlimULA_radfix_Rx}%
	}%
	\caption{$A_\mathrm{Rx}$ for a ULA in frontfire with a fixed array size of $40\lambda$.}%
	\label{fig:rxtxlimULAradfixRx}
\end{figure}%
\begin{figure}[t]
	\centering
	\ifbool{pdffigures}{%
	    \includegraphics{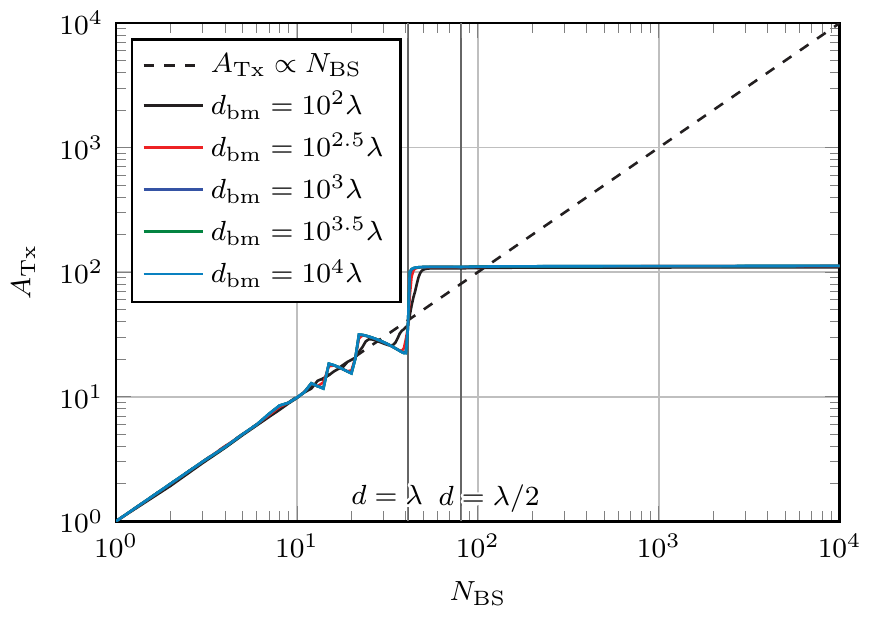}%
    }{%
	    \tikzsetnextfilename{rxtxlimULAradfixTx}%
    	\input{rxtxlimULA_radfix_Tx}%
    }%
	\caption{$A_\mathrm{Tx}$ for a ULA in frontfire with a fixed array size of $40\lambda$.}%
	\label{fig:rxtxlimULAradfixTx}
\end{figure}%
\begin{figure}[t]
	\centering
	\ifbool{pdffigures}{%
	    \includegraphics{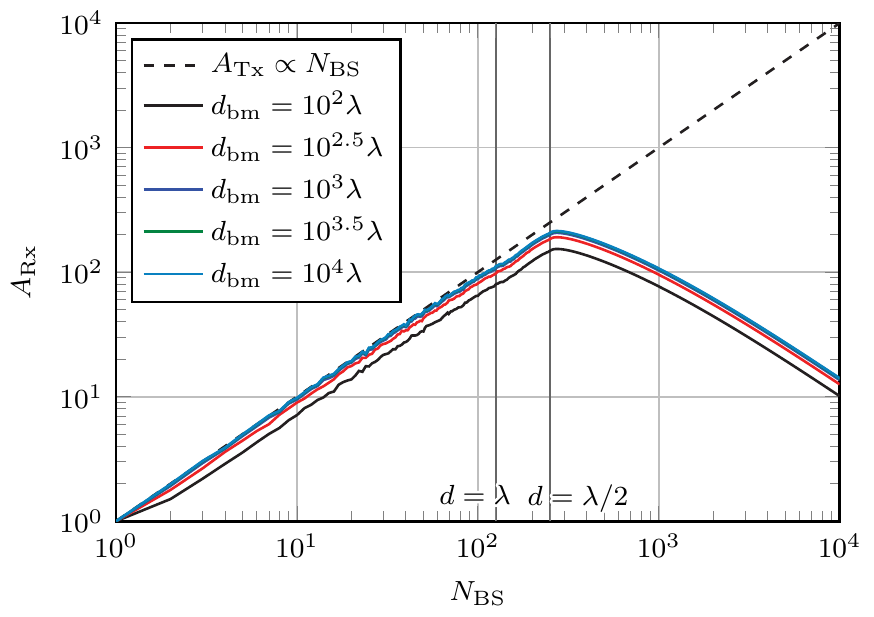}%
    }{%
	    \tikzsetnextfilename{rxtxlimUCAradfixRx}%
    	\input{rxtxlimUCA_radfix_Rx}%
    }%
	\caption{$A_\mathrm{Rx}$ for a UCA with fixed radius $r=20\lambda$.}%
	\label{fig:rxtxlimUCAradfixRx}
\end{figure}%
\begin{figure}[t]
	\centering
	\ifbool{pdffigures}{%
	    \includegraphics{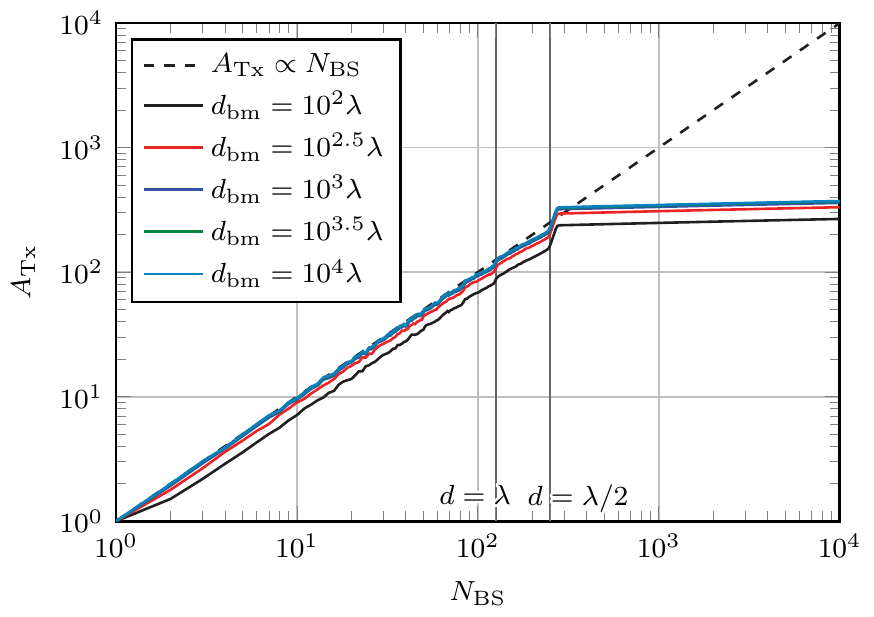}%
    }{%
	    \tikzsetnextfilename{rxtxlimUCAradfixTx}%
	    \input{rxtxlimUCA_radfix_Tx}%
	}%
	\caption{$A_\mathrm{Tx}$ for a UCA with fixed radius $r=20\lambda$.}%
	\label{fig:rxtxlimUCAradfixTx}
\end{figure}%
\begin{figure}[!t]
	\centering
	\ifbool{pdffigures}{%
	    \includegraphics{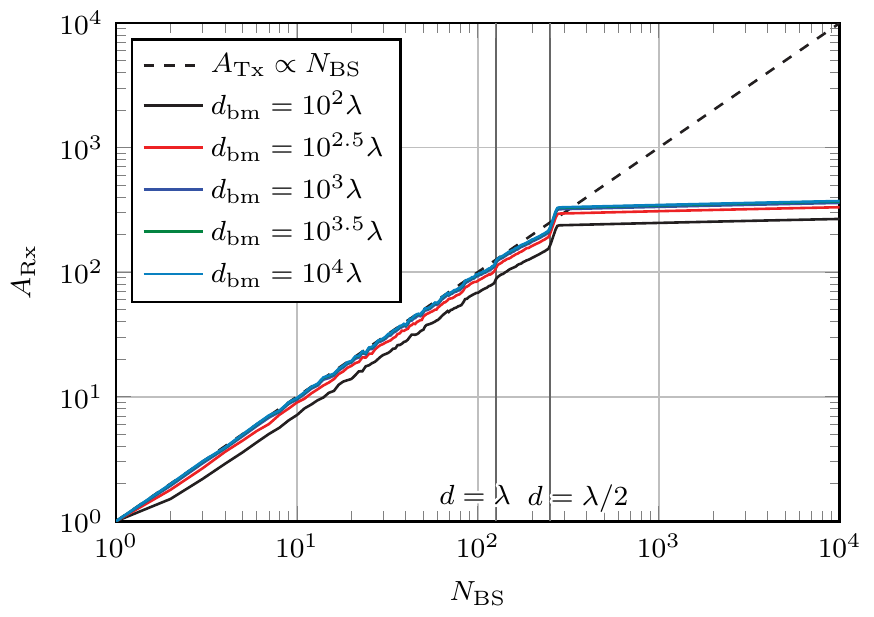}%
    }{%
    	\tikzsetnextfilename{rxtxlimUCAradfixthnoiseRx}%
	    \input{rxtxlimUCA_radfixthnoise_Rx}%
	}%
	\caption{$A_\mathrm{Rx}$ for a UCA with fixed radius $r=20\lambda$, antenna noise only.}%
	\label{fig:rxtxlimUCAradfixRxthnoise}
\end{figure}%
Consider a ULA with array size $40\lambda$ and a UCA with radius $r=20 \lambda$, see Fig.~\ref{fig:scenarios}. As $N_\mathrm{BS}$ increases, the antenna separation $d$ decreases. $A_\mathrm{Rx}$ and $A_\mathrm{Tx}$ are shown in Figs.~\ref{fig:rxtxlimULAradfixRx} to \ref{fig:rxtxlimUCAradfixTx} for different distances between the base station and the mobile. The curves for the ULA show that even for values slightly greater than $N_\mathrm{BS}=10$ antennas, $A_\mathrm{Rx}$ and $A_\mathrm{Tx}$ deviate from the linear increase expected from conventionally modeled systems, see \eqref{equ:convgain}. For the UCA, for $d> 1.26 \lambda$ (corresponding to $N_\mathrm{BS} \le 100$), \eqref{equ:convgain} holds approximately. Technically there is a small gap to $N_\mathrm{BS}$ in the array gain when the receiver is close to the base station, see the curve for $d_\mathrm{bm}=100\lambda$. 

However for smaller distances between the antennas, the trend for $A_\mathrm{Rx}$ and $A_\mathrm{Tx}$ is different. Considering the former, for the ULA there is a maximum at about 71 antennas ($d\approx 0.571\lambda$) and for the UCA at about 275 antennas ($d\approx 0.457\lambda$), and if the number of base station antennas is increased further, $A_\mathrm{Rx}$ decreases. Considering $A_\mathrm{Tx}$, it (almost) saturates at about 68 antennas for the ULA and 275 antennas for the UCA. If the number of base station antennas is increased further, $A_\mathrm{Tx}$ only increases slightly. Notably, the number of antennas for which the maximum $A_\mathrm{Rx}$ is obtained, and for which $A_\mathrm{Tx}$ starts to saturate, is (almost) independent of the distance to the mobile. There is a saturation, as the achievable array gain for a certain array size is bounded. The sharp increase in array gain for the ULA at about $d=\lambda$ can be explained by an increasing number of antennas, and a decreasing antenna separation at the same time, compared to~\cite{IvrlacNossekTowardaTheory}. 

In Fig.~\ref{fig:rxtxlimUCAradfixRxthnoise}, the receive array gain for the UCA is shown for the case when there is only the noise of the antennas, i.e., $\sigma_u = \SI{0}{\volt}, \sigma_i = \SI{0}{\ampere}$. In this case, the curves for $A_\mathrm{Rx}$ are (almost) the same as the curves for $A_\mathrm{Tx}$ in Fig.~\ref{fig:rxtxlimUCAradfixTx}, i.e., the coupling of the noise of the LNA and other analog components between the receive chain causes the decrease in $A_\mathrm{Rx}$.

\section{Numerical Results for the Array Gain in the Far Field}
\begin{figure*}[!t]
	\centering
	\ifbool{pdffigures}{%
		\includegraphics{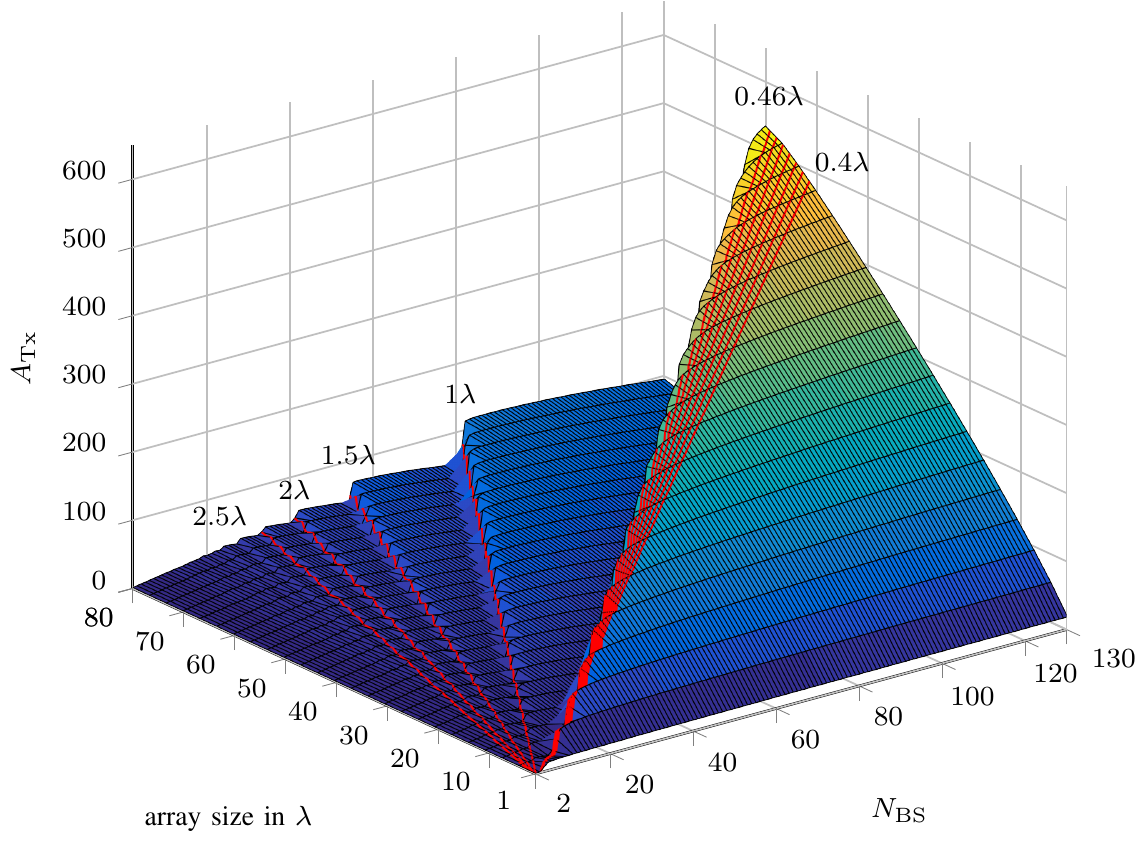}%
	}{%
	    \tikzsetnextfilename{3dantULAendfire}%
    	\input{3dantULA_endfirepng}%
    }%
	\caption{$A_\mathrm{Tx}$ for a ULA transmitting into endfire direction ($\varphi=90^\circ$), where the lines for $d=0.4\lambda$ to $0.46\lambda$ are in $0.01\lambda$ increments.}%
	\label{fig:ULAtxendfire}
\end{figure*}%
\begin{figure*}[!t]
	\centering
	\ifbool{pdffigures}{%
	    \includegraphics{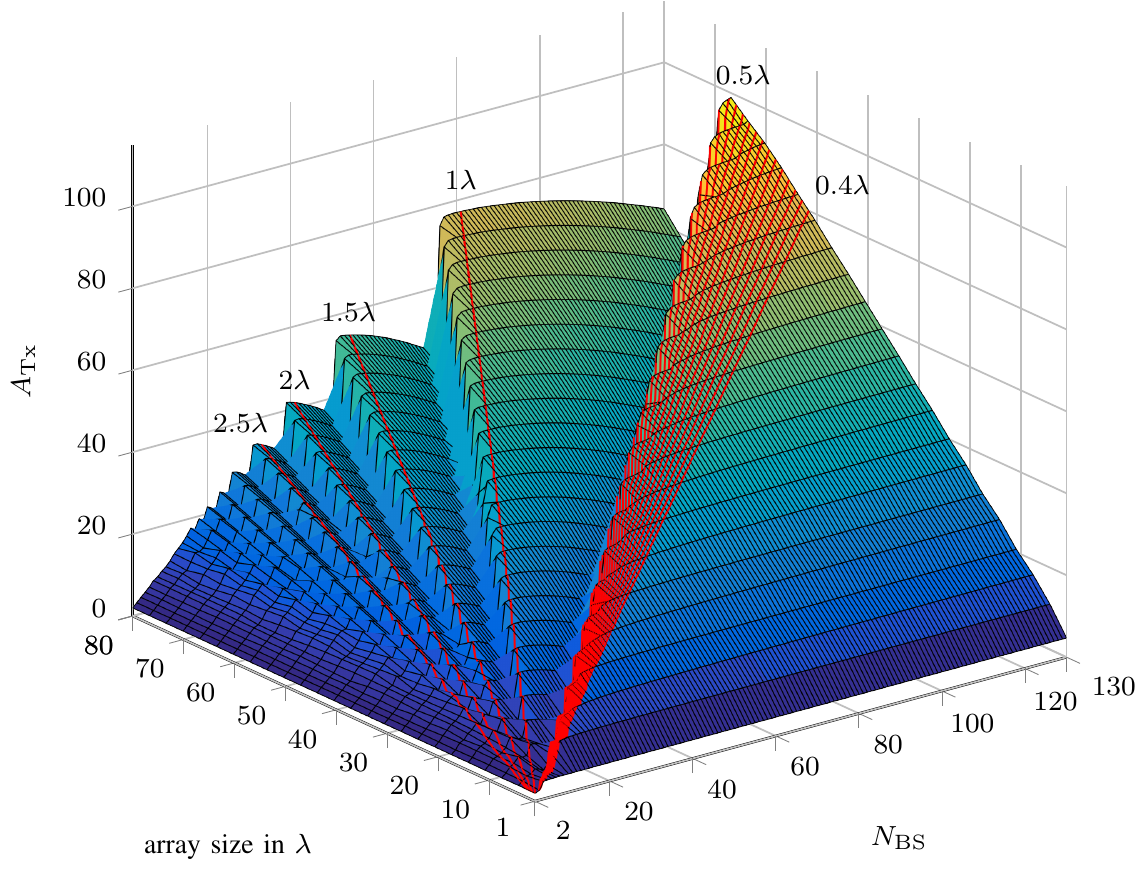}%
    }{%
    	\tikzsetnextfilename{3dantULA60deg}%
	    \input{3dantULA_60degpng}%
    }%
	\caption{$A_\mathrm{Tx}$ for a ULA transmitting into $\varphi = 60^\circ$ direction, where the lines for $d=0.4\lambda$ to $0.5\lambda$ are in $0.01\lambda$ increments.}%
	\label{fig:ULAtx60degfire}%
\end{figure*}
\begin{figure*}[!t]
	\centering
	\ifbool{pdffigures}{%
		\includegraphics{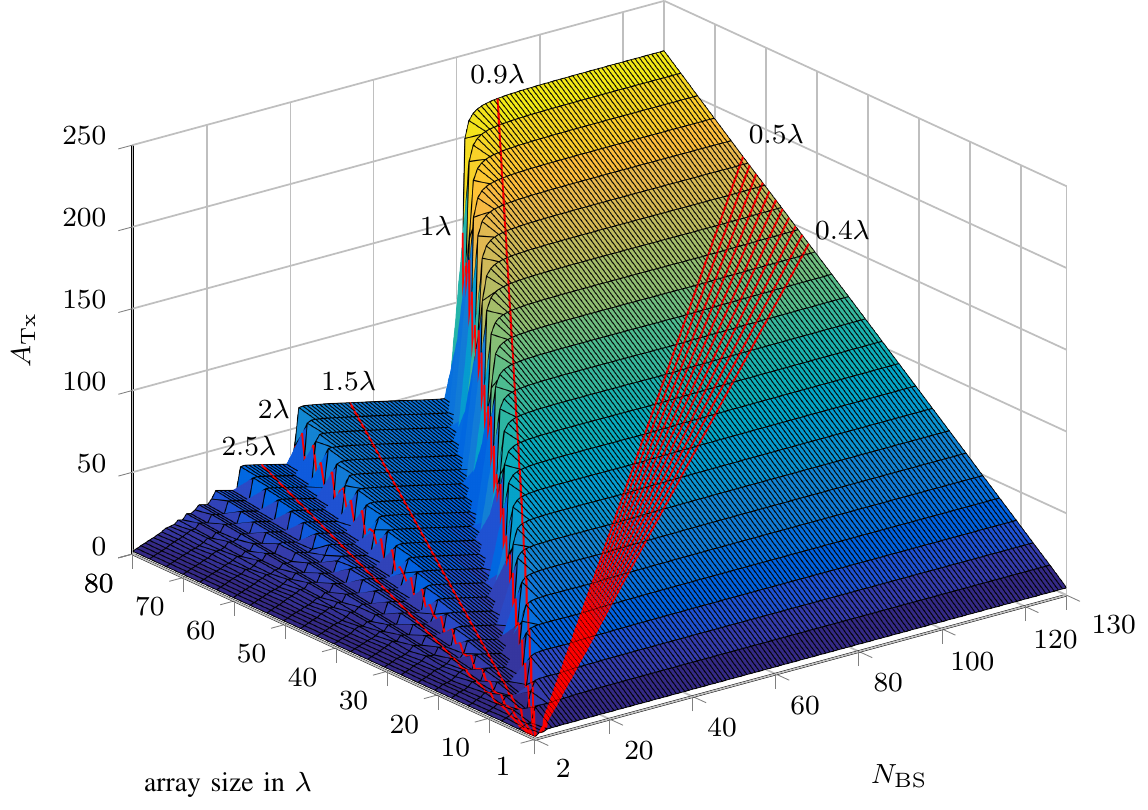}%
	}{%
		\tikzsetnextfilename{3dantULAfrontfire}%
		\input{3dantULA_frontfirepng}%
	}%
	\caption{$A_\mathrm{Tx}$ for a ULA transmitting into frontfire direction ($\varphi=0^\circ$), where the lines for $d=0.4\lambda$ to $0.5\lambda$ are in $0.01\lambda$ increments.}%
	\label{fig:ULAtxfrontfire}
\end{figure*}

%
In this section, we consider the transmit array gain in the far field for $\theta=\pi/2$, see \eqref{equ:atxff}, for ULAs and UCAs with different sizes and numbers of antennas $N_\mathrm{BS}$.
Firstly, consider the ULA in three scenarios: beamforming into the endfire ($\varphi=90^\circ$), into the $\varphi=60^\circ$ and into the frontfire direction ($\varphi=0^\circ$), see Figs.~\ref{fig:ULAtxendfire} to \ref{fig:ULAtxfrontfire}. The cut for frontfire direction with an array size of $40\lambda$ corresponds to Fig.~\ref{fig:rxtxlimULAradfixTx} with $d_\mathrm{bm}\to\infty\lambda$.
We can see that $A_\mathrm{Tx}$ depends on which directions the beamforming vector points to. Consider a fixed antenna array size. Then for the frontfire direction, saturation starts slightly below $d=\lambda$, but for the $60^\circ$ direction, saturation only starts slightly below $d=0.54\lambda$ and for the endfire direction saturation starts slightly below $d=\lambda/2$.  Further, we can observe that the larger the $N_\mathrm{BS}$, the closer the saturation starts to $d=\lambda, d\approx 0.54\lambda$ and $d=\lambda/2$ respectively. To maximize $A_\mathrm{Tx}$ the array should be positioned so that its endfire direction points to the angle of interest. If this is not possible, to optimize $A_\mathrm{Tx}$ for angles $-\varphi_0 \le \varphi \le \varphi_0$, where $\varphi_0$ is the maximum angle of interest that the array is transmitting to, $N_\mathrm{BS}$ should be chosen such that $d$ is at a peak for $\varphi=\varphi_0$. The maximum peak is slightly below $d=\lambda/2$ to $d=\lambda$ depending on $\varphi_0$ -- unless the loss factor $\gamma$ is too large. 
On the one hand, if $N_\mathrm{BS}$ is chosen to be slightly smaller, $A_\mathrm{Tx}$ falls into a valley for $\abs{\varphi}$ close to $\varphi_0$. On the other hand, $N_\mathrm{BS}$ is chosen to be slightly larger, either $A_\mathrm{Tx}$ only increases slightly for $d<\lambda/2$ and $A_\mathrm{Tx}$ becomes more sensitive to tolerances in the position of the antennas in the array, or $A_\mathrm{Tx}$ decreases again ($d>\lambda/2$).

\begin{figure*}[!t]
	\centering
	\ifbool{pdffigures}{%
		\includegraphics{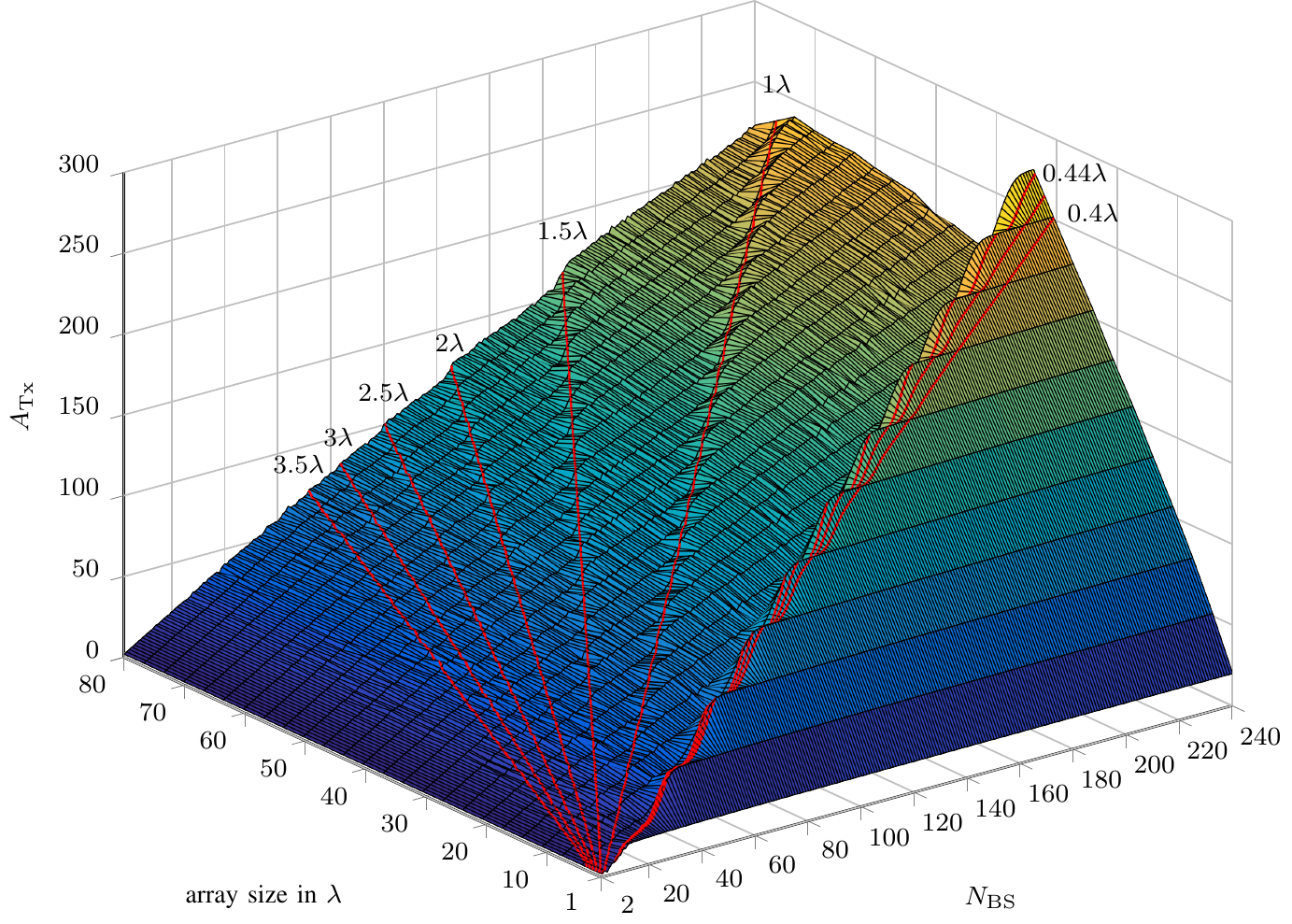}%
	}{%
		\tikzsetnextfilename{3dantUCAazimuthplane}%
		\input{3dantUCA_azimuthplanepng}%
	}%
	\caption{$A_\mathrm{Tx}$ for a UCA, where the lines for $d=0.4\lambda$ to $0.44\lambda$ are in $0.02\lambda$ increments.}%
	\label{fig:UCAtx}%
\end{figure*}

Secondly, consider a UCA with beamforming into a direction that lies on the line between an antenna and the center of the array (w.l.o.g. $\varphi = 0$), see Fig.~\ref{fig:UCAtx}. Here, similarly the cut for an array size of $40\lambda$ corresponds to Fig.~\ref{fig:rxtxlimUCAradfixTx} with $d_\mathrm{bm}\to\infty\lambda$. Similarly to the ULA transmitting into the endfire direction, $A_\mathrm{Tx}$ starts to saturate close to $d=\lambda/2$. The UCA behaves differently to the ULA. For transmission into the plane of the UCA, $A_\mathrm{Tx}$ is independent of $\varphi$ for an odd $N_\mathrm{BS}$, while for an even $N_\mathrm{BS}$ it only oscillates slightly with $\varphi$. Therefore, for maximum $A_\mathrm{Tx}$, $N_\mathrm{BS}$ should be chosen such that $A_\mathrm{Tx}$ is at the largest peak, i.e., the peak near $d=\lambda/2$, unless $\gamma$ is too large; then the peak of $A_\mathrm{Tx}$ near $d=\lambda$ is the largest.

According to the expectation from conventionally modeled systems, \eqref{equ:convgain} should hold, but for the ULA there are significantly larger maximum transmit array gains for front- and especially endfire ($A_\mathrm{Tx}\approx 219$ and $533$ for $N_\mathrm{BS}=130$), but also smaller maximum transmit array gains as seen for $\varphi=60^\circ$ ($A_\mathrm{Tx}\approx 99$ for $N_\mathrm{BS}=130$). For the UCA, there is not such a large direction-dependent variation, but the maximum is $A_\mathrm{Tx}\approx 152$ for $N_\mathrm{BS}=130$.

These figures could be used for further evaluations, e.g., whether using a single UCA for a base station deployment rather than non-cooperating ULAs in the typical 3 sectors leads to higher performance, with the same number of antennas in both scenarios.

\section{Conclusions}
Firstly, we have shown for two different massive MIMO scenarios, one with a fixed distance between the antennas and one with a fixed size of the array, the transmit array gain saturates above a certain number of antennas -- contrary to the expectation of a linear increase with the number of antennas when mutual coupling is neglected. Similarly, the receive array gain saturates when there is only thermal noise from the antennas at the receiver. However, if there is noise from the LNA and other components in the receive chains and no decoupling network, as it would be almost impossible to implement for massive MIMO, the coupling of this noise between the receive chains leads to a maximum of the receive gain for a certain number of antennas, and a decreasing receive gain for a larger number of antennas. 
Therefore it is essential to take mutual coupling into account in the design of large scale massive MIMO systems because otherwise, large array gains, which are unphysical, may be predicted. Secondly, we have derived practical design recommendations for ULAs and UCAs: ULAs should be oriented such that their endfire direction points in the direction of interest, and for both ULAs and UCAs, the optimum antenna spacing is slightly below $\lambda/2$. 
Future work includes the analysis of the array gain in a rich scattering environment.

\section*{Acknowledgment}
The authors would like to thank their colleague S. Bazzi for useful discussions and critical review of the paper.

\IEEEtriggeratref{15}
\bibliography{transactions1_bibtexnew}

\end{document}